 \documentclass[12pt,preprint]{aastex}

\usepackage{natbib}
\newcommand{\xmmnewton}{{\it XMM-Newton}}

\newcommand{\asca}{{\it ASCA}}
\newcommand{\einstein}{{\it Einstein}}
\newcommand{\chandra}{{\it Chandra}}
\newcommand{\fuse}{{\it FUSE}}
\newcommand{\hst}{{\it HST}}

\newcommand{\NH}{\mbox {$N_{\rm H}$}}
\newcommand{\nh}{\mbox {$N_{\rm H}$}}
\newcommand{\htwo}{\mbox {${\rm H}_{2}$}}
\newcommand{\hi}{H\,{\sc i}}

\newcommand{\cii}{C\,{\sc ii}}
\newcommand{\oiii}{O\,{\sc iii}}
\newcommand{\ovi}{O\,{\sc vi}}
\newcommand{\ovii}{O\,{\sc vii}}
\newcommand{\oviii}{O\,{\sc viii}}
\newcommand{\neix}{Ne\,{\sc ix}}
\newcommand{\nex}{Ne\,{\sc x}}
\newcommand{\mgxi}{Mg\,{\sc xi}}
\newcommand{\mgxii}{Mg\,{\sc xii}}
\newcommand{\sixiii}{Si\,{\sc xiii}}
\newcommand{\sixiv}{Si\,{\sc xiv}}

\newcommand{\about}{$\sim$\kern.03em}

\slugcomment{Accepted}

\shorttitle{Reverse Shock in the SNR 1E\,0102.2--7219}
\shortauthors{Sasaki et al.}

\begin{document}


\title{FUV and X-ray Observations of the Reverse Shock \\
in the SMC SNR 1E\,0102.2--7219}


\author{Manami Sasaki\altaffilmark{1}} 
\email{msasaki@cfa.harvard.edu}

\author{Terrance J. Gaetz\altaffilmark{1}}

\author{William P. Blair\altaffilmark{2}}

\author{Richard J. Edgar\altaffilmark{1}}

\author{Jon A. Morse\altaffilmark{3}}

\author{Paul P. Plucinsky\altaffilmark{1}}

\and

\author{Randall K. Smith\altaffilmark{2,4}}

\altaffiltext{1}{Harvard-Smithsonian Center for Astrophysics,
    60 Garden Street, Cambridge, MA 02138}

\altaffiltext{2}{Department of Physics and Astronomy, 
The Johns Hopkins University, 3400 North Charles Street, Baltimore, MD 21218}

\altaffiltext{3}{Department of Physics and Astronomy, 
Arizona State University, Box 871504, Tempe, AZ 85287-1504}

\altaffiltext{4}{NASA Goddard Space Flight Center, Code 662, Greenbelt, MD
20771}


\begin{abstract}
We present Far Ultraviolet Spectroscopic 
Explorer (\fuse) and X-ray Multi-mirror Mission (\xmmnewton) data for the
reverse shock of the O-rich supernova remnant (SNR) 
1E\,0102.2--7219 in the Small Magellanic
Cloud (SMC). The \fuse\ observations cover three regions with significantly
different optical [\oiii] intensities, all associated with the relatively 
bright part of the X-ray ring. 
Emission lines of \ovi\ $\lambda\lambda$1032, 1038 
are clearly detected in the \fuse\ spectra.  
By combining this \ovi\ doublet emission with the \ovii\ triplet and 
\oviii\ Ly$\alpha$ fluxes from the \xmmnewton\ spectra and assuming
a non-equilibrium ionization (NEI) model 
with a single ionization timescale
for the spectra, we are able to
find a narrow range of temperatures and ionization timescales that are 
consistent with the respective line ratios. 
The increase of the ionization timescale $\tau$ from North 
($\tau \approx 0.6 \times 10^{11}$~s~cm$^{-3}$) to South-East 
($\tau \approx 2 \times 10^{11}$~s~cm$^{-3}$) is
indicative of increasing density in the X-ray bright ring, 
in good agreement with the optical [\oiii] emission which is strongest in the 
South-East. 
However, if we assume a plane-parallel shock model with a distribution of
ionization timescales, the \ovi\ emission appears to be inconsistent with
\ovii\ and \oviii\ in X-rays. 

We also analyze the total \xmmnewton\ EPIC-MOS 1/2 spectra for the three
regions. The X-ray spectra are dominated by strong emission lines of O, Ne, 
and Mg, however, we detect an emission component that accounts for 14 -- 25\% 
of the flux and can be attributed to shocked ISM.  
We find that there is no consistent set
of values for the temperature and ionization timescale which can explain
the observed line ratios for O, Ne, and Mg.  This would be consistent with
a structured distribution of the ejecta as the O, Ne, Mg would have
interacted with the reverse shock at different times.
\end{abstract}


\keywords{Shock waves --- supernova remnants --- ultraviolet: ISM
--- X-rays: individual (1E\,0102.2--7219)}


\section{Introduction}

\begin{figure*}
\centering
\caption{\label{images}
\chandra\ ACIS-S3 image (left) and \hst\ WFPC2 [\oiii] image (right)
with \fuse\ MDRS aperture positions. See 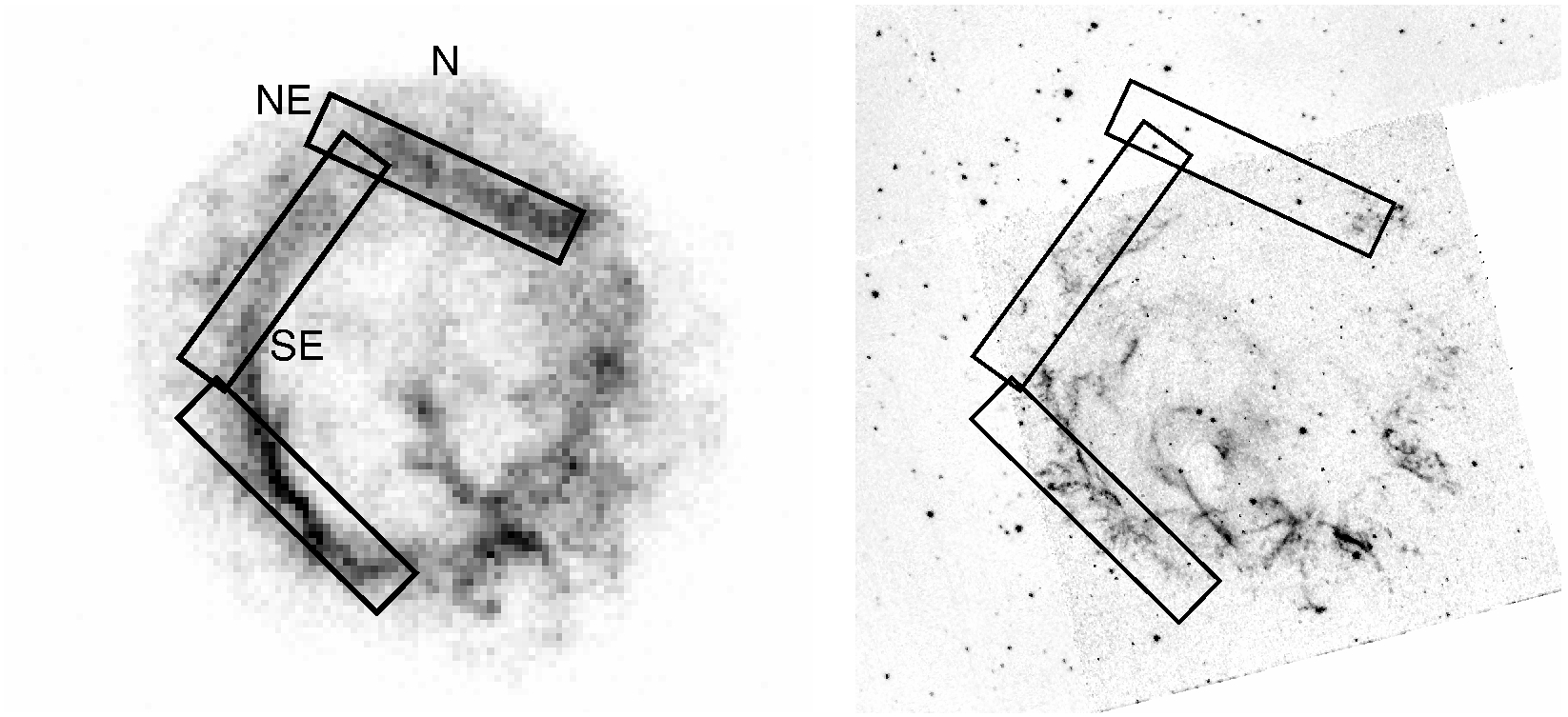.
}
\end{figure*}

The supernova remnant (SNR) 1E\,0102.2--7219 (hereafter E0102) 
is the brightest X-ray SNR in the Small Magellanic Cloud (SMC) and was
discovered with \einstein\ \citep{1981ApJ...243..736S}.
In X-rays, it has a ring-like appearance and its emission is dominated
by H-like and He-like ions of O and Ne
\citep[e.g.,][]{2000ApJ...534L..47G,2000ApJ...543L..61H,2001A&A...365L.231R,2001A&A...365L.237S,2004ApJ...605..230F}.
It consists of a bright X-ray ring and a much fainter plateau with a sharp
outer edge that surrounds the ring (Fig.\,\ref{images}). 
This bright X-ray ring is 
caused by the reverse shock of the SNR, while the outer boundary of the
faint plateau marks the location of the blast wave. The diameter of the
bright ring is \about28\arcsec, and the diameter of the outer rim is
\about44\arcsec.
The radio emission in the 6\,cm wavelength band \citep{1993ApJ...411..761A} 
is located outside the bright X-ray emission but within the outer faint X-ray 
rim \citep{2000ApJ...534L..47G}.

In the optical band, there is diffuse H$\alpha$ emission surrounding the 
remnant, but a distinctive H$\alpha$ `hole' at the location of the SNR.
A filamentary shell of optical [\oiii] emission with a diameter of
\about30\arcsec\ is seen \citep{1981ApJ...248L.105D}, which is
presumably emission from the ejecta. Further outwards,
the remnant is surrounded by a faint, diffuse, partial shell of [\oiii] with 
a diameter of \about60\arcsec\ that is thought to be the result of 
photoionization by the precursor. Emission lines of O, Ne, C, and 
Mg have been detected in the UV and optical spectra \citep{1981ApJ...248L.105D,
1998RMxAC...7...21M,2000ApJ...537..667B}, but H, He, and Fe are notably absent.
\citet{1983ApJ...268L..11T} estimated an age of \about1000~yrs for the
remnant based on the optical diameter and the line of sight velocity range.
Based on abundance estimates from optical and UV data, 
\citet{2000ApJ...537..667B} suggested that the progenitor was possibly a 
massive O star that, after becoming a Wolf-Rayet star, exploded as a Type Ib 
supernova.
More recently, \citet{2005ApJ...619..839C} argued that E0102 is the remnant of 
a Type IIL/b supernova with a progenitor that lost most of the H envelope 
before the explosion. 

We present new observations with the {\it Far Ultraviolet Spectroscopic 
Explorer} (\fuse) to study the \ovi\ emission in regions in the X-ray bright 
ring with differing distributions of [\oiii] and X-rays. In addition, we 
analyze X-ray observations with \xmmnewton\ and \chandra\ and compare the O 
emission in the Far-UV (FUV) to the X-ray emission.

\section{Observations and Reductions}

\subsection{\fuse\ Data}

The \fuse\ satellite was launched in June 1999; details of its 
operation and on-orbit performance are provided in the papers by 
\citet{2000ApJ...538L...1M} and \citet{2000ApJ...538L...7S}. 
The \fuse\ instruments cover
the 905 -- 1187~\AA\ band with a nominal point source spectral resolution
of $R \approx 20,000$. The \fuse\ telescope has four optical channels,
each with its own mirror and grating. Two mirrors and gratings are 
coated with silicon carbide (SiC channels), and the other two mirrors 
and gratings are coated with Al and a lithium fluoride overcoat (LiF 
channels). One SiC/LiF channel pair is sensed by one microchannel 
plate intensified detector, and the other channel pair is sensed by 
a second detector.  The SiC channels provide wavelength coverage from 
\about905~\AA\ to 1100~\AA, whereas the LiF channels cover the wavelength 
range of \about1000~\AA\ to 1187~\AA.
Each of the two detectors is divided into two segments (1A and 1B, 
2A and 2B), so that 8 spectra are obtained for each pointing (LiF1A/B, SiC1A/B, 
LiF2A/B, SiC2A/B). 

At the focal plane of each channel is an assembly holding 
three spectrograph apertures, dubbed LWRS, MDRS, and HIRS. In time-tag
data mode, spectra through all three apertures are recorded simultaneously,
but only one aperture is designated as primary.  The primary science 
apertures for our program are the 4\arcsec\ $\times$ 20\arcsec\ MDRS 
apertures.

After launch, small thermally-induced distortions were discovered that
cause misalignments of these apertures.  While periodic alignment
activities are performed to keep the channels approximately aligned,
motions of 3 -- 8\arcsec\ occur on an orbital timescale ($\sim$100
minutes), preventing detailed co-alignment during a given integration.
The LiF1 channel was used for guiding in the data presented here,
so that the location of the source is fixed for this channel.
Due to the drift, not all the counts from the source
fall on the detector. As this is not taken into account when the flux of the
calibrated spectrum is computed, the flux of the SiC and LiF2 channels become
lower than the real flux which is measured in the LiF1 channel. The spectra 
of the SiC and LiF2 channels can be corrected by scaling the continuum flux to 
the correct continuum flux as determined in the LiF1 spectrum.

We observed three positions around the bright X-ray ring of E0102. 
The positions were selected to have differing distributions of 
[\oiii] in the optical and \ovii, \oviii\ in the X-ray (Table \ref{obstab}). 
The observations were scheduled at times where the aperture position 
angles could be tailored to the positions being observed.
Figure \ref{images} shows the projected positions on a \chandra\ image and 
on an \hst\ WFPC2 [\oiii] image. The net \fuse\ integration times are 
12.4~ksec for the north pointing (N), 15.6~ksec for the northeast pointing 
(NE), and 8.2 + 5.2~ksec for the southeast pointing (SE, two visits).
We also performed a dedicated off-source position with MDRS for 
estimating the background.  However, the spectra from the background 
pointing show more significant contamination by background starlight
than the on-source positions.  Therefore, the background observation 
has not been subtracted from the source spectra.

The data have been processed by the CalFUSE 3.0 pipeline, and the 
resulting spectra are shown in Figure \ref{fusespec}.
The only emission lines detected from the SNR are the broadened 
\ovi\ doublet lines.
We have modeled these spectra using functions available in IDL and routines 
from an IDL library including the fitting function MPFIT\footnote{\it 
http://astrog.physics.wisc.edu/\about craigm/idl/idl.html} that performs
least-squares curve fitting.  Line fluxes and derived surface
brightnesses in \ovi\ are provided in Table \ref{oneovi} (see 
Sect.\,\ref{ovidoublet} for more details).

\subsection{X-ray Data}

There are numerous public archival X-ray data sets for E0102 obtained with 
\chandra\ \citep{1996SPIE.2805....2W} and \xmmnewton\
\citep{2001A&A...365L...1J}. We first considered using 
the Advanced CCD Imaging Spectrometer (ACIS) data from the \chandra\ archive. 
\chandra\ data have far superior spatial resolution in comparison with
\xmmnewton.
However, we have determined that the brightest portions of E0102 as observed 
with the ACIS-S array in full frame mode are affected by pile-up, while the 
ACIS-S sub-array data and ACIS-I array data with the remnant on-axis have poor 
photon statistics. 
Pile-up causes two or more photons to be detected as a single event. As
the observed energy is the sum of the energies of two or more photons, the
energy spectrum is thus distorted. 
Compared to \chandra\ observations, \xmmnewton\ observations are less affected
by pile-up due to shorter frame-times and lower spatial resolution (thus
smearing emission from bright sources over a larger region of the detector).
\xmmnewton\ is equipped with three European Photon Imaging Cameras (EPIC);
one is a pn-type CCD (EPIC-PN) and two are MOS-type CCDs 
(EPIC-MOS1/2). Both the ACIS and the EPICs are sensitive in the energy band 
of \about0.3 -- 10.0~keV.
For SNRs, we are in general interested in soft X-rays ($\lesssim\ 4.0$~keV).
Because the sensitivity of \xmmnewton\ in the energy range of 0.5 -- 2.0~keV 
is higher and the spectral resolution of EPIC-MOS detectors is better than
that of ACIS, we decided to use \xmmnewton\ data primarily, except for the 
examination of the faint outer rim. The outer rim is much fainter than the
bright ring and has much less pile-up. \chandra's spatial resolution helps to 
isolate the rim emission from the bright ring emission.

\begin{figure}[t]
\centering
\includegraphics[width=0.45\textwidth,clip]{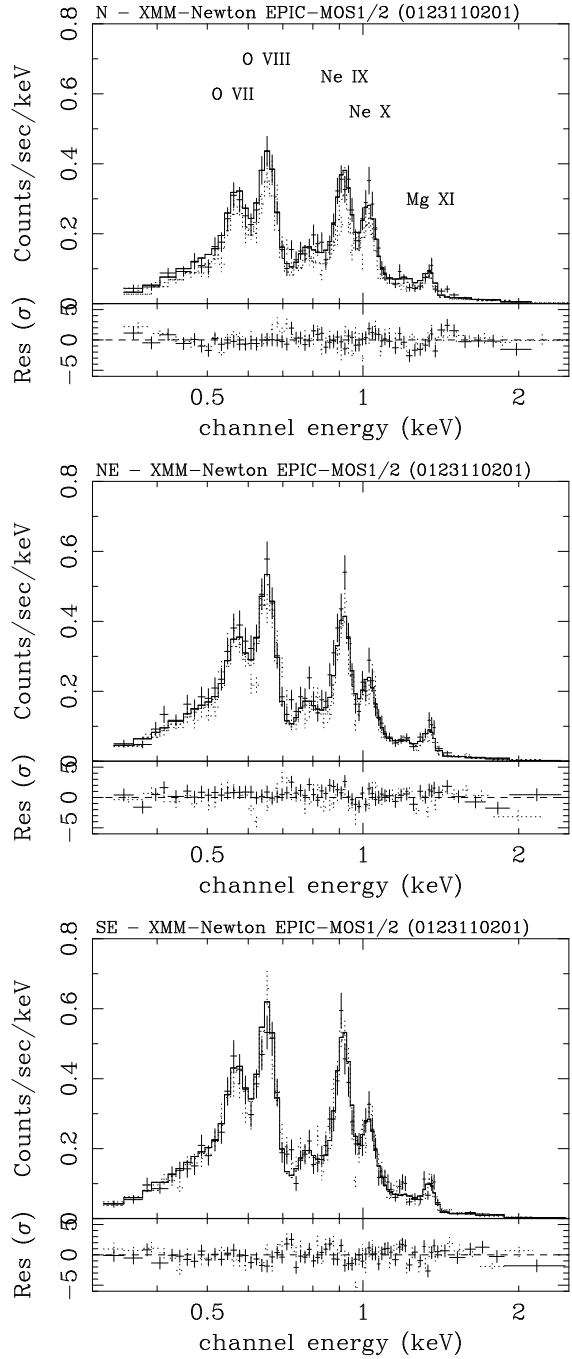} 
\caption{\label{xmmspec} 
\xmmnewton\ EPIC-MOS1 (solid) and MOS2 (dashed) spectra of the observation 
0123110201 for the regions N, NE, and SE.
Fit model includes three VNEI spectral components (see Sect.\,\ref{vneifit}). 
The prominent O, Ne, and Mg lines are labeled in the N spectrum.
}
\end{figure}

We selected archival data of three \xmmnewton\ observations which were taken
early in the mission (April 2000 and April 2001): 
ObsIDs 0123110201, 0123110301, and 0135720601. Because we want to study the 
emission lines, we use EPIC-MOS1 and MOS2 data which have better spectral 
resolution than EPIC-PN. 
The net exposure times after removing times with soft proton flares 
are \about16~ks, 13~ks, and 20~ks for the
observations 0123110201, 0123110301, 0135720601, respectively.
The data were obtained in small window mode resulting in a short frame time
(0.3~s; for comparison, frame time of the \chandra\ ACIS full frame mode is 
3.2~s) and negligible pile-up. The data are processed with XMMSAS 6.1.0.
We extracted spectra of the reverse shock at the positions of the \fuse\ 
apertures (Fig.\,\ref{xmmspec}). The brightest lines in the X-ray spectra at 
these positions are the \ovii\ triplet lines and the \oviii\ Ly$\alpha$ line. 

\section{Analysis}

\subsection{\ovi\ Doublet}\label{ovidoublet}

Although the diffuse \ovi\ emission lines are broad and faint, these line 
features are well detected in the \fuse\ LiF1A and LiF2B
spectra. The SiC1A and SiC2B spectra also cover the spectral range around the 
\ovi\ lines, however, the statistics of those data are poor due to the lower 
effective areas. Also, the SiC channels are more affected by the misalignment 
of the channels than the LiF2B channel and the emission from the brighter 
parts of the remnant may have largely moved out of the aperture.
We see significant \ovi\ emission in both LiF1A and LiF2B data. The
rescaling of the LiF2B spectra by effective area and continuum flux
shows that the LiF2B seems to have mainly covered the same \ovi\ emission
as the guiding LiF1A channel. Therefore, we combine the LiF1A and LiF2B spectra 
for each pointing and analyze the co-added spectra 
around the \ovi\ lines (1020 to 1050\AA). The strongest emission lines in the 
spectra of the E0102 pointings are terrestrial airglow lines. In the wavelength 
band around the \ovi\ doublet $\lambda\lambda$1032, 1038 
(1031.926\AA, 1037.617\AA, respectively), there is a strong \hi\ airglow line 
at 1025.7\AA, while the \ovi\ emission is relatively faint, in part because of
the broadening of the lines.
We use this \hi\ airglow line to calibrate the wavelength and to
confirm that the spectral resolution is consistent with the aperture size.
The \ovi\ doublet fluxes are derived by modeling the emission features between 
\about1030\AA\ and 1040\AA.

Interstellar absorption is significant in the Far-UV and needs be taken 
into account when determining the line fluxes.
We include absorption lines of \htwo\ and \cii\, and 
correct the flux for reddening through extinction.
The \htwo\ absorption has been derived using {\tt H2ools} developed 
by \citet{2003PASP..115..651M}.
For the column density of \htwo\ we use the Milky Way (MW) and SMC values from 
\citet{2002ApJ...566..857T}, i.e., median values from 13 stars in the 
northern part of the SMC. 
There is an evident absorption feature at \about1036\AA\ which is a
\cii\ $\lambda\lambda$1036.3367, 1037.0182 doublet. We model this feature
as a line at \about1036 \AA\ with free width. 
We also model 
absorption by \ovi\ along the line of sight,
including both the MW and the SMC components.

\begin{figure}[t]
\centering
\includegraphics[bb=17 15 450 1073,angle=0,width=0.45\textwidth,clip]{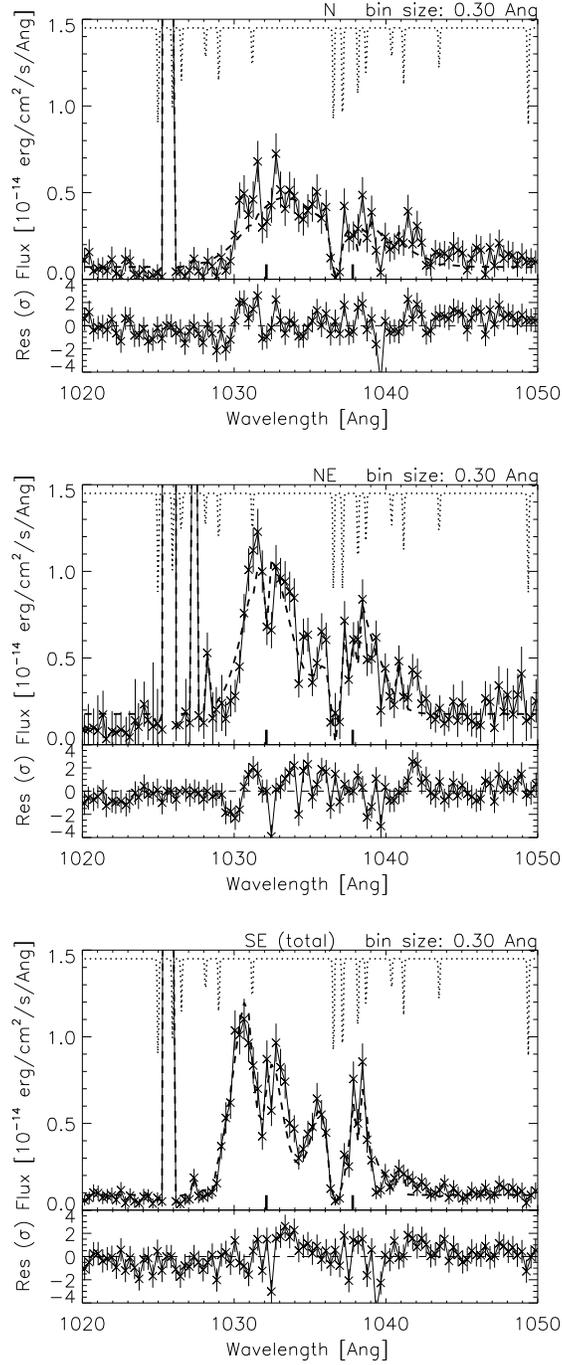} 
\caption{\label{fusespec} 
Close-ups of \fuse\ spectra of N, NE, and combined SE pointings. 
LiF1A and LiF2B spectra are combined. The thick dashed line shows the model
(see Sect.\,\ref{ovidoublet}).
The thick lines at the bottom mark the rest wavelengths of the \ovi\ lines.
The dotted line shows the \htwo\ absorption in an arbitrary scale.
}
\end{figure}

We combine night and day data (instead of using only night data to reduce the
airglow emission) because of the poor photon statistics.
Nevertheless, the sensitivity of \fuse\ allows the broad \ovi\
emission lines from this faint diffuse source to be well detected 
(see Fig.\,\ref{fusespec}).
The spectra are binned with 0.3 \AA\ bins (corresponding to
\about87~km~s$^{-1}$).
Using the fitting routine MPFIT, we fit the \ovi\ $\lambda\lambda$1032, 1038 
lines with two broad Gaussians, and the continuum is modeled as a constant. 
The widths of \ovi\ $\lambda\lambda$1032, 
1038 Gaussian lines are tied together, and the relative positions of the lines
are fixed to the wavelength separation: i.e.\ we fit the \ovi\ $\lambda$1032 
line (free wavelength) and tie the \ovi\ $\lambda$1038 line to the
\ovi\ $\lambda$1032 wavelength + 5.691\AA.
We also considered fitting each line with many zero-width Gaussians which 
fill the observed width of the broad lines, to see if we can find different 
velocity components in the lines. The zero-width Gaussians
form pairs with fixed separation between the \ovi\ $\lambda$1032 and the 
\ovi\ $\lambda$1038 components. However, the fits did not improve significantly.
The normalization of each line is a free parameter except for the N pointing 
data, which have much poorer statistics. In the case of the N data, the 
normalization of the $\lambda$1038 line is fixed at half of the $\lambda$1032 
line normalization (corresponding to the optically thin limit).
From the line energy and width, we derive a velocity shift and FWHM.
The flux is corrected for extinction, using the Galactic extinction
curve of \citet{1990ApJS...72..163F} and
\citet{1999PASP..111...63F} as well as extinction curves of the Magellanic 
Clouds from \citet{2003ApJ...594..279G}.
To obtain the de-reddened flux, we use $E(B-V)$ = 0.08, assuming that
the extinction is caused half by the SMC and half by the MW, as suggested by 
\citet{1989ApJ...338..812B,2000ApJ...537..667B}.
We compute the 90\% confidence errors for the line energy, width, and flux.
The FWHM, shifts, and fluxes are listed in Table \ref{oneovi}. 
Figure \ref{fusespec} shows the spectra around $\lambda$1036 with the fits 
overplotted and fit residuals.

The \ovi\ emission is faintest in the N pointing and brightest in the SE 
pointing. In both the NE and SE pointings, the fits show that the gas has a 
velocity dispersion of about 900~km~s$^{-1}$, while the faint \ovi\ line 
feature in the N pointing spectrum has a FWHM of $1540\pm350$~km~s$^{-1}$. 
This is most likely not thermal line broadening, but instead indicates a
velocity structure in the remnant (see Sect.\,\ref{specresult}). The rest SMC 
velocity is \about165~km~s$^{-1}$ \citep{2002ApJS..139...81D}, so
the broad \ovi\ emission lines are blue-shifted by 
\about100~km~s$^{-1}$ in the NE pointing and by
\about330~km~s$^{-1}$ in the SE pointing.
However, in the N pointing, the broad \ovi\ emission line feature is 
red-shifted by 
\about220~km~s$^{-1}$. This might indicate a non-symmetric expansion of 
the remnant.

\subsection{X-ray Emission Lines}\label{lindiagn}

To combine the X-ray data with the FUV data, we perform an X-ray emission line 
analysis using the H-like and He-like lines of O, Ne, and Mg. 
The X-ray spectra have been binned with a minimum of 20 counts per bin 
to allow the use of $\chi^2$ statistics. We fit the
EPIC-MOS1/2 spectra using the X-ray spectral fitting package XSPEC. The model
consists of Gaussian lines at the known line energies of 
O, Ne, Mg, and Si (i.e.\ 
\ovii\ triplet at \about0.57~keV, \oviii\ Ly$\alpha$ 
at 0.654~keV, \neix\ triplet at \about0.91~keV, \nex\ Ly$\alpha$ at 1.022~keV, 
\mgxi\ triplet at \about1.34~keV, \mgxii\ Ly$\alpha$ at 1.473~keV,
\sixiii\ triplet at \about1.84~keV, \sixiv\ Ly$\alpha$ at 2.006~keV),
a continuum-only model using the Astrophysical Plasma Emission Code 
\citep[APEC,][]{2001ApJ...556L..91S}, and line-of-sight absorption.
The foreground absorption model consists of a component for the foreground MW 
absorption with a fixed \nh\ of $5.36 \times 10^{20}$~cm$^{-2}$ 
\citep{1990ARA&A..28..215D} and a component with SMC abundances 
\citep{1992ApJ...384..508R} to be fit.
All the Ly$\alpha$ lines are well fit with zero-width Gaussians.
The CCD resolution does not allow the components of the triplet lines 
(\ovii, \neix, and \mgxi) to be resolved, but at least two zero-width 
Gaussians are necessary to reproduce the line features. Therefore, we fit the 
triplet emission with two Gaussians at the energies of the forbidden and the
resonance lines which are the two stronger lines and span the energy range
of the triplet. 
The forbidden and the resonance lines are fit separately, 
i.e., their flux ratio is free in the fits, 
and the flux of the intercombination line is included in the fluxes of the
fitted lines. 
We use the sum of the fluxes as the flux of the triplet. 


Oxygen lines are the strongest lines in both the \fuse\ and the \xmmnewton\ 
spectra of E0102. We compare line fluxes obtained from \fuse\ (\ovi\ doublet) 
and EPIC-MOS (\ovii\ triplet and \oviii\ Ly$\alpha$) to line fluxes predicted
by models.
The observed line ratios are listed in Table \ref{ratiotab}.
For the comparison with the observed lines, we use line fluxes computed with 
{\tt neiline}\footnote{\it 
http://cxc.harvard.edu/cont-soft/software/NEIline.1.00.html},
which evaluates line emissivities for an
ionizing or recombining plasma.
Line fluxes are calculated by evolving the plasma at a given initial 
temperature for a given ionization timescale.
As this calculation assumes a single ionization timescale for the plasma, 
we also model line fluxes using a plane-parallel shock model with a single 
temperature and a distribution of ionization timescales appropriate for a 
plane-parallel shock \citep[VPSHOCK model in XSPEC; 
see][]{2001ApJ...548..820B}. We assume SMC abundances for all 
elements except for O, Ne, and Mg for both models. For the elements O, Ne, and 
Mg with enhanced abundances we use the numbers obtained from the global fits 
of the X-ray spectrum (see Sect.\,\ref{vneifit}).

The diagrams in Figure \ref{linediagn} show the line ratios 
\ovii\ triplet/\oviii\ Ly$\alpha$,
\ovi\ doublet/\ovii\ triplet, and \ovi\ doublet/\oviii\ Ly$\alpha$ 
derived from \fuse\ and \xmmnewton\ spectra and the error 
estimates based on the 90\% confidence intervals for the line fluxes as a 
function of the ionization timescale $\tau = n_{\rm e}t$ and temperature $kT$
for the two models.
In the VPSHOCK model, the parameter $\tau$ is the upper limit of the
ionization timescale that has a distribution characteristic for a
plasma shocked by a steady plane-parallel shock.
For the three line ratios for O, we evaluate confidence regions for the ratios
as a function of $kT$ and $\tau$. 
The lines in black show the ratio between \ovii\ and \oviii\ observed in 
X-rays (\ovii\ triplet/\oviii\ Ly$\alpha$). The cyan and orange lines compare 
\ovi\ in Far-UV with \ovii\ and \oviii\ ( \ovi\ doublet/\ovii\ triplet
and \ovi\ doublet/\oviii\ Ly$\alpha$).
The results show that in the case of the {\tt neiline} model, the confidence 
regions for the ratios \ovii\ triplet/\oviii\ Ly$\alpha$, \ovi\ doublet/\ovii\ 
triplet, and \ovi\ doublet/\oviii\ Ly$\alpha$ intersect, yielding a 
well-defined $kT$ vs.\ $\tau = n_{\rm e}t$ region for each aperture:
for N, $kT \approx 0.3 - 0.4$~keV and 
$\tau \approx (0.5 - 2) \times 10^{11}$~s~cm$^{-3}$,
for NE, $kT \approx 0.3 - 0.4$~keV and $\tau > 0.4 \times
10^{11}$~s~cm$^{-3}$, and 
for SE, $kT \approx 0.25 - 0.35$~keV and $\tau > 0.7 \times
10^{11}$~s~cm$^{-3}$.
While the $kT$-$\tau$ regions for NE and SE overlap, the result for N
indicates higher $kT$ and lower $\tau$.
However, the diagrams for the plane-parallel shock model show no overlap
between the confidence regions of the line ratios.

\begin{figure*}[t]
\centering
\includegraphics[width=0.95\textwidth,clip]{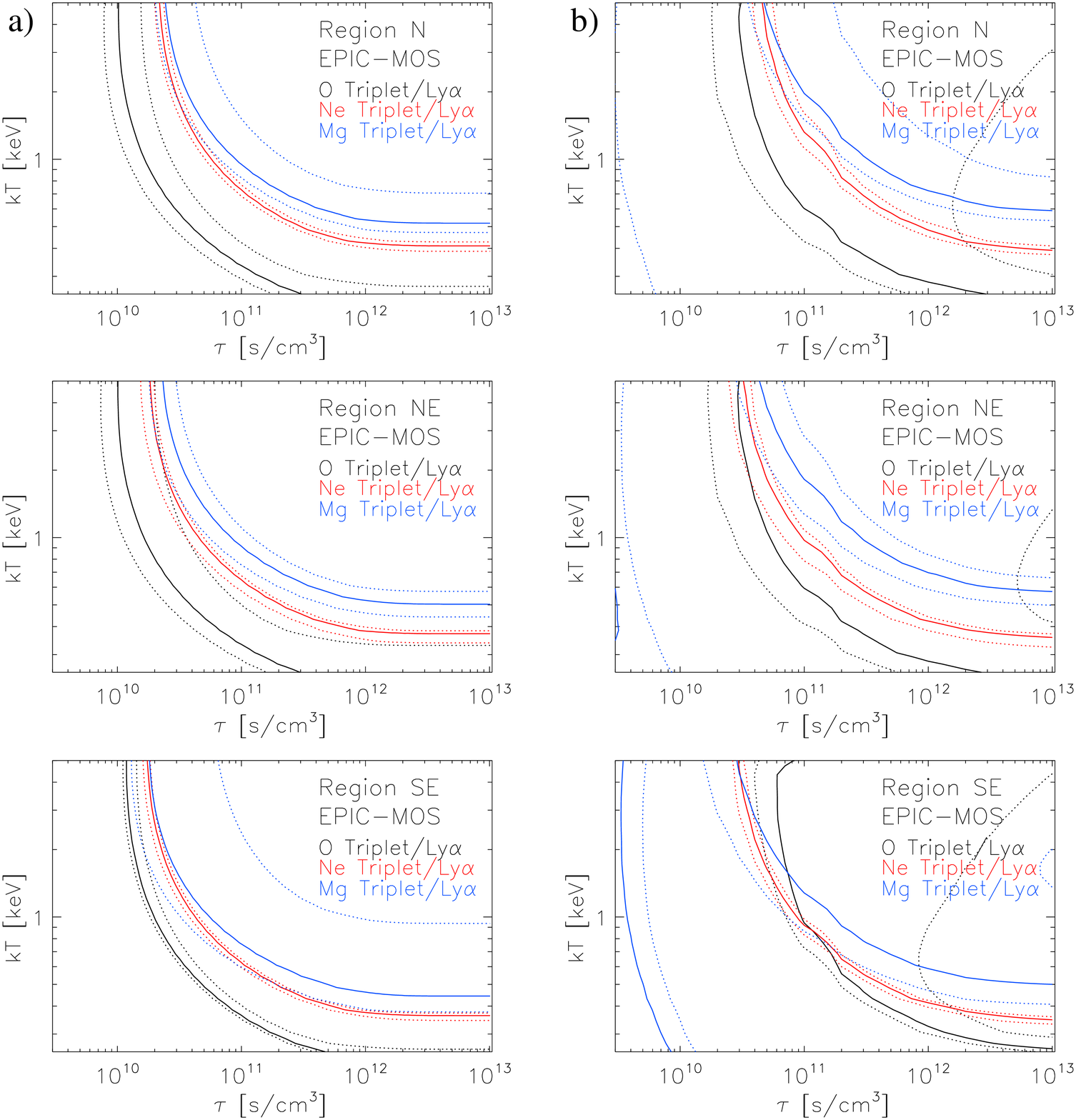} 
\caption{\label{tripladiagn} 
Line ratios \ovii\ triplet/\oviii\ Ly$\alpha$, \neix\ triplet/\nex\ Ly$\alpha$, 
and \mgxi\ triplet/\mgxii\ Ly$\alpha$ for the three regions corresponding to the
apertures of the \fuse\ pointings N, NE, and SE obtained from EPIC-MOS spectra
(solid line, 90\% errors as dotted lines). 
The used models are a) {\tt neiline} and b) VPSHOCK (in XSPEC).
The parameter $\tau$ of the VPSHOCK model is the upper limit of the
ionization timescale.
}
\end{figure*}

We also computed and compared the line ratios of 
\ovii\ triplet/\oviii\ Ly$\alpha$, \neix\ triplet/\nex\ Ly$\alpha$, 
and \mgxi\ triplet/\mgxii\ Ly$\alpha$ to check whether different elements 
might indicate different plasma conditions. The results are shown in 
Figure \ref{tripladiagn}. 
As the diagrams show, for {\tt neiline}, the confidence regions for O, Ne, and 
Mg have little overlap and there are no conditions where all three intersect.
There seems to be no single set of temperatures and ionization timescales that
simultaneously describes the O, Ne, and Mg ratios. This is consistent with
previous results. In analyzing \asca\ data for E0102, 
\citet{1994PASJ...46L.121H} have found it necessary to introduce separate 
plasma components for each element. Analysis of \chandra\ High Energy 
Transmission Grating Spectrometer (HETGS) data has shown that the dispersed 
line emission images have different radii for lines of different elements, 
suggesting progressive ionization by the reverse shock propagating through 
the ejecta \citep{2004ApJ...605..230F}.
Another explanation for the lack of overlap between the ions is that
the X-ray emission arises from a distribution of shocks with different
velocities as suggested for SNR 1987A by \citet{2005ApJ...628L.127Z}.
In the case of a plane-parallel shock model (VPSHOCK) the three confidence
regions overlap for $kT \gtrsim\ 1$~keV and $\tau\ \lesssim\ 
10^{11}$~s~cm$^{-3}$. 
However, when the \ovi\ line results from \fuse\ are included, 
there is no region of common temperatures and ionization timescales which 
satisfies the O line ratios. This might be indicating that the bulk of the
\ovi\ emission arises from a different location with different plasma 
conditions than the location from which the bulk of the \ovii\ and \oviii\
emission arises.

\subsection{Global Fitting of the X-ray Spectrum}\label{vneifit}

To analyze the global X-ray spectra of the three regions, we first fit the 
spectra in XSPEC using the VNEI model with NEI version 2.0 for a plasma 
in non-equilibrium ionization (NEI) with variable abundances. 
This model assumes a single ionization timescale for the emitting plasma
and is a simplification compared to the plane-parallel shock model
\citep[see][]{2001ApJ...548..820B}.
The NEI version 2.0 models use a newer atomic database \citep[Astrophysical 
Plasma Emission Database, APED,][]{2001ASPC..247..161S} than the versions 
1.0 and 1.1 and reproduces more accurate line emissivities. However, they do 
not contain inner-shell processes.
For the absorption we include two components: one component 
represents the foreground MW absorption with a fixed \nh\ of 
$5.36 \times 10^{20}$~cm$^{-2}$ and the other represents the SMC absorption.
For each region, the EPIC-MOS1 and MOS2 spectra for the three
observations are fit simultaneously with a single model.

Although we fit the spectra with various sets of initial parameters,
models with a single VNEI component fail to fit the spectra adequately. In all 
cases, the model predicts a lower flux than the observed spectrum next to 
the \mgxi\ triplet line feature, exactly where the \mgxii\ Ly$\alpha$ line 
would be expected. 
Because the fits for the X-ray bright ring are dominated by the strong \ovii, 
\oviii, \neix, and \nex\ lines, we conclude that the elements O, Ne, and Mg 
might not be ionized to the same extent, with Mg being more highly excited
than the other elements.

We therefore try a model with additional VNEI components to represent
different plasma components. 
First, we include a component for the ISM shocked by the blast wave: 
To estimate this component, we use \chandra\ ACIS-S3 data (ObsIDs 
\dataset [ADS/Sa.CXO#obs/05123] {5123} and 
\dataset [ADS/Sa.CXO#obs/05130] {5130}) 
and extract a spectrum in the X-ray plateau covering the outer blast
wave, i.e.\ a ring centered on RA = 01$^{h}$~04$^{m}$~02.0$^{s}$~, 
Dec = --72\degr\ 01\arcmin\ 53.2\arcsec\,
with inner radius of 19\arcsec\ and outer radius of 23\arcsec\,.
This emission is very faint and not affected by pile-up. In the \xmmnewton\
EPIC-MOS1/2 data it is not possible to separate clearly the emission from the 
blast wave and the ejecta because of the poor spatial resolution of \xmmnewton. 
Fitting the blast wave spectrum extracted from \chandra\ data with one VNEI 
component (absorption component as above), we obtain:
$kT = 1.1$~keV, $\tau = 5.7 \times 10^{10}$~s~cm$^{-3}$. Therefore, we fit 
the \xmmnewton\ spectra of parts of the bright X-ray ring with a model
including a VNEI component for the ISM with $kT$ and $\tau$ fixed, and
all the abundances set to the SMC ISM abundances \citep{1992ApJ...384..508R}.
For the ejecta emission, we include two VNEI components: 
a component for the emission dominated by O, fixing the parameters for
$kT$ and $\tau$ to the results from the line diagnostics 
(see Fig.\,\ref{linediagn}) with all other element abundances set to zero, and
a component accounting for the rest of the ejecta emission, consisting of 
Ne and Mg (with abundances for the other elements set to zero), with free $kT$ 
and $\tau$. 
The same set of absorbing columns (Galactic and SMC as described 
above) is used for both components. The fits are better than with a single 
VNEI model, and the fit parameters are listed in Table \ref{xmmpar}.
The spectra of the ObsID 0123110201 data and the best fit model, obtained 
by a simultaneous fit of all spectra of the three ObsIDs, 
are shown in Figure \ref{xmmspec}.

\section{Discussion}

\subsection{Results from the Spectral Analysis}\label{specresult}

If the line broadening in the \fuse\ spectra were due to thermal motion, the
FWHM of about 900~km~s$^{-1}$ would correspond to an unreasonably high 
temperature of 3$\times 10^{8}$~K. The analysis of the EPIC-MOS data shows that 
the plasma temperature is about 10$^{7}$~K.
There are at least three scenarios that can explain this discrepancy.
First, it is possible that the collisionless heating of the electrons behind 
the reverse shock front is negligible. 
\citet{2000ApJ...543L..61H} argue that this is the case at the outer blast
wave and that significant particle acceleration is required there.
In such a case, the oxygen ions, and thus also \ovi, have a much high 
temperature than the electrons which are responsible for the X-ray emission.
Second, the electron temperature might have been underestimated by applying 
a single ionization timescale model (VNEI). 
Third, the broadening might be caused by turbulence or by bulk motions
rather than by thermal motions. 
\citet{2004ApJ...605..230F} derive expansion velocities of about 
$\pm$1000~km~s$^{-1}$ from \chandra\ HETGS data for the bright X-ray 
ring. 
The velocities measured from the HETGS data could be explained by geometric 
effects, if E0102 has, e.g., a tilted, expanding ring geometry or a distorted 
shell with density variations. \citet{2004ApJ...605..230F} suggest that the 
difference in velocity shifts in different parts of the remnant might 
indicate that the southeastern part of the 
remnant shell is approaching and the northwestern part is receding. This is 
consistent with the velocity shifts obtained for the broad \ovi\ emission 
lines which show blue-shifts in the NE and SE pointings, whereas the lines in 
the N pointing indicate a red-shift (Sect.\,\ref{ovidoublet}).

The X-ray emission of the bright X-ray ring is mainly ejecta emission. Because
the spectra in all three regions are not well fit by a single 
component NEI model, we have fit the spectra with two NEI model components 
with variable abundances of O (first component) and Ne and Mg (second 
component) for the emission from the ejecta shocked by the reverse shock, and 
an additional NEI model to account for the emission from the ISM 
shocked by the blast wave. We calculate the unabsorbed flux (0.3 -- 10.0~keV) 
of the ejecta and shocked ISM for each region and compare the flux fraction: 
The ratio between the surface brightness of the total ejecta emission and the 
ISM emission, $S_{\rm ejecta}/S_{\rm ISM}$, is 6, 3, and 3 for regions N, NE, 
and SE, respectively. The surface brightness of the ISM component is 
$S_{\rm ISM} \approx 5 \times 10^{-15}$~erg~cm$^{-2}$~s$^{-1}$~arcsec$^{-2}$
in all three regions. For comparison, the surface brightness of the emission of
the shocked ISM in the outer rim is $S_{\rm outer rim} = 4.7 \times
10^{-15}$~erg~cm$^{-2}$~s$^{-1}$~arcsec$^{-2}$, in good agreement with the
estimated ISM component for the emission of the X-ray bright ring at the
positions of the \fuse\ pointings. The differences between the total flux of the
three regions in the X-ray bright ring seem to be caused by differences in the 
emission from the ejecta shocked by the reverse shock as well as by variations
in absorption as obtained from the spectral fits.

\begin{figure*}[t]
\centering
\includegraphics[width=0.95\textwidth,clip]{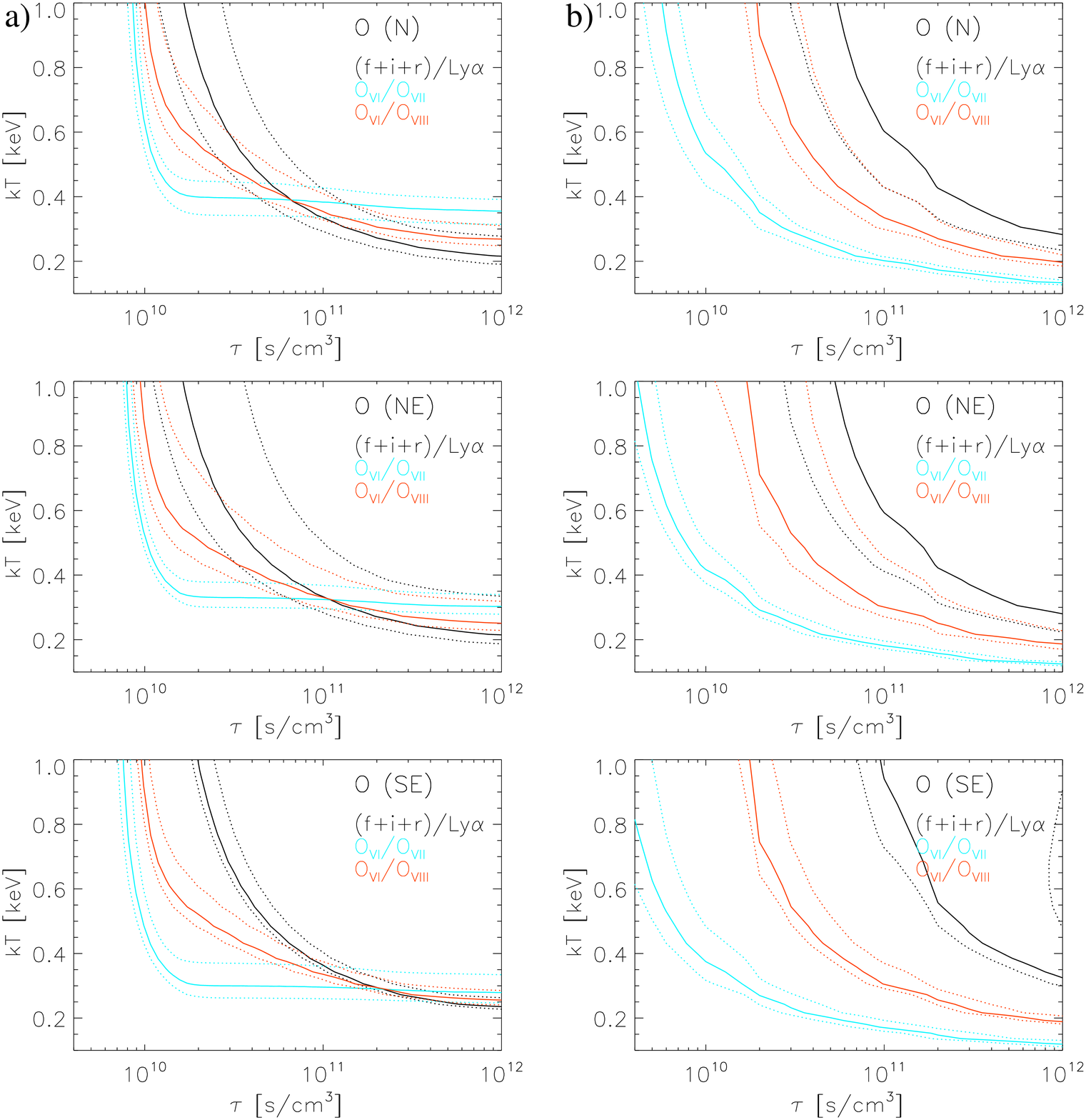} 
\caption{\label{linediagn} 
Line ratios \ovii\ triplet/\oviii\ Ly$\alpha$, \ovi\ doublet/\ovii\ triplet, 
and \ovi\ doublet/\oviii\ Ly$\alpha$ for the regions corresponding to the
apertures of the \fuse\ pointings N, NE, and SE (solid line). The estimated 
range for the error is shown with dotted lines. 
To calculate the line ratios, we used the models a) {\tt neiline} and 
b) VPSHOCK (in XSPEC). In the VPSHOCK model, $\tau$ is the upper limit of the
ionization timescale that has a distribution typical for a plane-parallel
shock.
}
\end{figure*}

The line ratio diagrams in Figure \ref{linediagn}a show that the \ovi\ 
seen in FUV and the \ovii\ and \ovii\ X-ray emission are compatible with the 
assumption of an ionizing plasma with a single ionization timescale $\tau$, as 
calculated by {\tt neiline}.
In this case, the line diagnostics of the O emission obtained with \fuse\ and 
\xmmnewton\ show that the ionization timescale of the O gas differs in the 
three regions, increasing from N to SE:
$\tau \approx 6 \times 10^{10}$~s~cm$^{-3}$ in the N region,
$1 \times 10^{11}$~s~cm$^{-3}$ in the NE region, and
$2 \times 10^{11}$~s~cm$^{-3}$ in the SE region.
The temperatures in the three regions are comparable: $kT$ = 0.3 -- 0.5~keV.
This result may indicate that the density in the ejecta is higher in the
southeastern part of the X-ray bright ring, consistent with the optical data
which show that the [\oiii] emission is strongest in the SE. 
If we instead assume a plane-parallel shock with a distribution of
ionization timescales using the VPSHOCK model, we do not obtain a consistent 
set of $kT$ and the upper limit of $\tau$ for the \ovi, \ovii, and \oviii\ 
lines. However, the diagrams in Figure \ref{tripladiagn}b indicate that the 
flux ratios of emission lines of O, Ne, and Mg in X-ray, and thus also \ovii\ 
and \oviii, are consistent with a common temperature and ionization timescale 
assuming the plane-parallel shock model. 
As the diagrams in Figure \ref{linediagn}b compare 
\ovi\ in FUV with the oxygen emission in X-ray, the fact that \ovi, 
\ovii, and \oviii\ show no overlap might suggest that the oxygen line emission 
in the FUV and the oxygen line emission in X-rays together are not consistent 
with the assumed model. As copious [\oiii] 
emission is observed in the optical, there must also be denser shocked material 
and the \ovi\ emission might be related to it as well. Therefore, multiple 
shocked plasmas or a range of velocities are possibly involved in the FUV and 
X-ray emission of E0102. 

These results show that we find a few possible scenarios for the emission from 
this SNR. However, as we are limited by the size of the aperture of \fuse\ and 
the spatial resolution of \xmmnewton\, we are not able to constrain whether the 
emission arises from a single shock or multiple shocks. 
Moreover, the shock can be ionizing or recombining and thus radiative.
Although we have optical [\oiii] images and X-ray images with \ovii\ and
\oviii\ emission of high spatial resolution, we lack spatial information on
\ovi. If it were possible to obtain \ovi\ imaging data we would be able to 
compare the distribution of O in the remant and thus to determine if the shock
is ionizing or recombining. Furthermore, hydrodynamical models for reverse 
shocks are necessary to understand the distribution of elements as well as 
physical parameters like $kT$ or $\tau$ in relatively young SNRs like E0102.

\subsection{Mass Estimates}

To derive the ejecta mass, we numerically estimate the volume of the SNR 
included in the slit, assuming that the SNR emission is from a thick shell.
The outer radius of the shell is assumed to coincide with the outer edge of 
the bright X-ray ring: $R_{\rm s}$ = 16\farcs0$\pm$1\farcs0 = 4.7$\pm$0.3~pc
\citep[for $D = 60$~kpc,][]{1999IAUS..190..569V}; the inner radius of the 
bright X-ray ring is taken to be 
$r_{\rm s}$ = 12\farcs0$\pm$1\farcs0 = 3.5$\pm$0.3~pc.
The resulting volume, $V$, of the observed emitting gas is the intersection 
between this thick shell and a projected rectangle with the slit 
size ($A =$ 4\arcsec$\times$20\arcsec\ = 6.77~pc$^{2} = 6.45 \times
10^{37}$~cm$^{2}$). The results are listed in Table \ref{esttab}. We assume an
error of \about20\% for the estimated volumes, as there are systematic
uncertainties including the actual shape of the shell.

The \chandra\ HETGS spectra show that the plasma mainly consists of O, Ne, and 
Mg \citep{2004ApJ...605..230F}. The EPIC-MOS fits also show that the ejecta 
emission results from O, Ne, and Mg. Therefore, we assume a metal-rich plasma 
consisting of O, Ne, and Mg excited to H-like, He-like, or fully ionized 
states.

%
%

The XSPEC VNEI emissivity normalization (Table \ref{xmmpar}) is
$K = 10^{-14}/(4 \pi D^{2}) \int n_{\rm e} n_{\rm H} dV$,
where $D$ is the distance to the source [cm], $n_{\rm H}$ is the hydrogen
number density [cm$^{-3}$], and $n_{\rm e}$ is the electron number density 
[cm$^{-3}$]. 
In the case of a plasma with high overabundance of elements other 
than hydrogen, $n_{\rm H}$ needs to be replaced by a reference density, $n_{\rm
ref}$, that is a sum over the number densities of the included elements,
thereby taking the abundances into account.
The electron density is $n_{\rm e,i} = \alpha_{\rm i} n_{\rm i}$, where 
$n_{\rm i}$ is the density of element i. The factor $\alpha_{\rm i}$ depends 
on the assumed mean charge state for each element; we assume 
$\alpha = 7, 9, 11$, for O, Ne, and Mg, respectively.
Assuming $D = 60$~kpc$= 1.85 \times 10^{23}$~cm and uniform density in the
thick shell, we estimate $n_{\rm O}$, $n_{\rm Ne}$, and 
$n_{\rm Mg}$ for the two components for each region (see Table \ref{esttab}).
Using these number densities, we obtain the O, Ne, and Mg mass for the entire 
remnant. 
We include a filling factor $f$ to account for the unknown 
volume fraction of the X-ray emitting plasma.
The densities are proportional to $f^{-0.5}$.
Assuming a thick shell with an outer radius of
$R_{\rm s}$ = 4.7$\pm$0.6~pc and an inner radius of 
$r_{\rm s}$ = 3.5$\pm$0.6~pc, we obtain $M_{\rm O} 
%
%
= 4\pm2~f^{0.5}~M_{\sun}$, $M_{\rm Ne} = 1.6\pm0.4~f^{0.5}~M_{\sun}$, 
$M_{\rm Mg} = 0.2\pm0.1~f^{0.5}~M_{\sun}$. If we assume a thick ring with 
$V = 6.6 \times 10^{57}$~cm$^{-3}$ as estimated by \citet{2004ApJ...605..230F}, 
we derive: $M_{\rm O}= 4\pm1~f^{0.5}~M_{\sun}$, $M_{\rm Ne} = 1.5\pm0.3~f^{0.5}~M_{\sun}$,
$M_{\rm Mg} = 0.2\pm0.1~f^{0.5}~M_{\sun}$.
We estimate a few $M_{\sun}$ in O, indicating a massive progenitor.

If we assume a volume fraction of unity, 
the oxygen mass $M_{\rm O}$ is consistent with the result from \chandra\ HETGS 
by \citet{2004ApJ...605..230F} who also assumed $f = 1$. 
By comparing their $M_{\rm O}$ estimate with 
the models of \citet{1997NuPhA.616...79N}, \citet{2004ApJ...605..230F} suggest
that the mass of the progenitor of E0102 was \about32~$M_{\sun}$. 
\citet{2000ApJ...537..667B} compare abundance ratios based on optical and UV
lines to values derived by \citet{1997NuPhA.616...79N} and conclude that the 
E0102 progenitor is best described by a 25~$M_{\sun}$ model. 
Our mass estimates for O and Mg as well as the mass ratios
agree well with the 25 -- 30~$M_{\sun}$ model 
\citep{1997NuPhA.616...79N}, while the Ne mass estimate is slightly 
higher than the model prediction. 
Even for a filling factor as low as $f \approx 0.1$, the mass estimates
are consistent with a massive progenitor \citep{1997NuPhA.616...79N}.

In these calculations, there are still some uncertainties that we have to be
aware of. As the reverse shock has not yet reached the center of the
remnant, there is still unshocked material interior to the 
reverse shock. If the unshocked material has the same 
abundance distribution as the shocked material (i.e., well-mixed ejecta) the 
ratios will not be much altered, but the estimated masses would be lower limits.
Furthermore, the ionization potential increases with the atomic number;
the shock might be strong enough to ionize O all the way up to the He-like and 
H-like stages, but not as efficient in ionizing Mg. Therefore, a part of 
Mg could be in lower ionization stages and not emitting in X-rays. 
Finally, assuming a too high volume filling factor can result in 
overestimating the total masses. Numerical calculations have shown that the 
ejecta shocked by the reverse shock form thin layers that can be deformed by 
Rayleigh-Taylor instabilities, decreasing the fraction of the observed ejecta 
in the apertures \citep{1992ApJ...392..118C,2001ApJ...560..244B}.

\section{Summary}

We have observed three regions in the bright X-ray ring of the O-rich 
SNR 1E\,0102.2--7219 in the SMC using the \fuse\ MDRS 
(4\arcsec$\times$20\arcsec) aperture.
These three regions cover parts of the remnant ring with varying
distributions of [\oiii] in the optical, and \ovii\ and \oviii\ in the X-ray. 
The ring is caused by the reverse shock propagating into the ejecta. In the 
northern region (N), the \fuse\ spectrum shows very faint but velocity 
broadened \ovi\ emission. In the northeastern (NE) and the southeastern (SE) 
parts of the SNR ring, there is significant \ovi\ emission with broad \ovi\
$\lambda\lambda$1032, 1038 lines, indicating a velocity dispersion of 
about 900~km~s$^{-1}$. The emission from the SE pointing appears to be 
blue-shifted by \about200~km~s$^{-1}$ relative to the emission from the 
NE pointing.

We have also analyzed the \xmmnewton\ EPIC-MOS1/2 spectra of E0102 in the
same regions as in the \fuse\ observations. 
For the combined analysis of the FUV and X-ray data, we consider two 
cases: 1.\ shocked plasma described by a single temperature and single 
ionization timescale (using the models {\tt neiline} and VNEI) and 2.\ 
plane-parallel shock with a distribution of ionization timescales (using the
VPSHOCK model).
In the first case, combining \ovi\ doublet emission 
from \fuse\ data with \ovii\ triplet and \oviii\ Ly$\alpha$ in EPIC-MOS1/2 
spectra yields $kT \approx 0.4$~keV and 
$\tau \approx 6 \times 10^{10}$~s~cm$^{-3}$ for the N region, 
$kT \approx 0.4$~keV and 
$\tau \approx 1 \times 10^{11}$~s~cm$^{-3}$ for the NE region,
and $kT \approx 0.3$~keV and 
$\tau \approx 2 \times 10^{11}$~s~cm$^{-3}$ for the SE region.
The ionization timescales of the O gas seem to increase towards the
south ($\tau_{\rm N} < \tau_{\rm NE} < \tau_{\rm SE}$), suggesting a 
higher density in the south. 
The total EPIC-MOS1/2 spectra are well fit with a three component 
non-equilibrium ionization model; two VNEI components are necessary to 
describe the ejecta emission (one for the O plasma, one for Ne + Mg plasma)
and an additional VNEI component is used to model the emission from the shocked 
ISM. The temperature of the Ne + Mg component is higher ($kT > 1$~keV) than 
the O component.
For the plane-parallel shock model, however, the observed ratios of 
\ovi/\ovii\ and \ovi/\oviii\ are not consistent with the \ovii/\oviii\ ratio,
in that there is not a common region of allowed temperatures and ionization
timescales.  When the triplet and Ly$\alpha$ ratios are considered for O, Ne, 
and Mg in the X-ray bandpass alone, the observed ratios do have significant
overlap in temperature ($kT >$ 1.0~keV) and ionization timescale ($\tau < 
1\times 10^{11}$~s~cm$^{-3}$).  These results might indicate that the 
\fuse\ and \xmmnewton\ data are sampling multiple shocks with different 
conditions.  To better understand these results in the future, we still require 
information on the spatial distribution of \ovi\ in FUV as well as hydrodynamic 
models describing the structure of the reverse shock.

From the X-ray emission, we estimate the mass of O, Ne, and Mg in the entire 
remnant. Making the assumption that the remnant is a thick shell, we 
obtain $M_{\rm O} = 4\pm2~f^{0.5}~M_{\sun}$, $M_{\rm Ne} = 1.6\pm0.4~f^{0.5}~M_{\sun}$,      
$M_{\rm Mg} = 0.2\pm0.1~f^{0.5}~M_{\sun}$. A comparison with previous measurements and
progenitor models shows that our results are consistent with a massive
progenitor.

\acknowledgments

We thank the anonymous referee for useful critiques and suggestions
which materially improved this paper.
We are grateful to Daniel Dewey, John Raymond, and Ravi Sankrit for helpful 
discussions.
This work is based on observations made with the NASA-CNES-CSA {\it
Far Ultraviolet Spectroscopic Explorer}. \fuse\ is operated for NASA by the 
Johns Hopkins University under NASA contract NAS5-32985.
The presented work is also based on observations obtained with \xmmnewton, an 
ESA science mission with instruments and contributions directly funded by 
ESA Member States and NASA.
The work was supported by the NASA/\fuse\ grant NAG5-12295, CXC 
contract NAS8-03060, and NASA grant GO1-2060X.
The \chandra\ X-ray Observatory Center is operated by the 
Smithsonian Astrophysical Observatory for and on behalf of the 
National Aeronautics Space Administration under contract NAS8-03060.



\clearpage


\clearpage

\clearpage

\clearpage 

\clearpage

\clearpage

\clearpage

\begin{deluxetable}{lccccc}
\tabletypesize{\footnotesize}
\tablecaption{\label{obstab} \fuse, \xmmnewton, and \chandra\ data used for the 
analysis.}
\tablehead{
\multicolumn{1}{c}{Obs.\ ID\tablenotemark{a}} & 
\multicolumn{2}{c}{Pointing direction} & 
\multicolumn{1}{c}{Aperture/} & 
\multicolumn{1}{c}{Start time (UT)} & 
\multicolumn{1}{c}{Exposure} \\
\multicolumn{1}{c}{} & 
\multicolumn{1}{c}{RA} & \multicolumn{1}{c}{Dec} & 
\multicolumn{1}{c}{Instrument} & \multicolumn{1}{c}{} & 
\multicolumn{1}{c}{night+day\tablenotemark{b}} \\
\multicolumn{1}{c}{} & \multicolumn{2}{c}{(J2000.0)} & 
\multicolumn{1}{c}{} & \multicolumn{1}{c}{} & \multicolumn{1}{c}{[ksec]} }
\startdata
\multicolumn{6}{c}{\fuse} \\
\noalign{\smallskip}\tableline\noalign{\smallskip}
C0830101 (N) \hspace{1.6cm} & 01:04:01.74 & --72:01:39.1 & MDRS & 2003-09-19 11:35:05 & 12 \\
C0830201 (NE) & 01:04:04.24 & --72:01:45.1 & MDRS & 2002-06-06 23:51:41 & 16 \\
C0830301 (SE1) & 01:04:04.04 & --72:02:01.9 & MDRS & 2003-08-29 07:54:23 & 8 \\
C0830302 (SE2) & 01:04:04.04 & --72:02:01.9 & MDRS & 2004-08-21 14:52:32 & 5 \\
\noalign{\smallskip}\tableline\noalign{\smallskip}
\multicolumn{6}{c}{\xmmnewton} \\
\noalign{\smallskip}\tableline\noalign{\smallskip}
0123110201 & 01:03:47.25 & --72:01:58.3 & EM1/2, SW\tablenotemark{c} & 2000-04-16 20:09:26 & 16 \\
0123110301 & 01:03:47.19 & --72:01:58.3 & EM1/2, SW\tablenotemark{c} & 2000-04-17 04:43:36 & 13\\
0135720601 & 01:03:45.79 & --72:00:51.1 & EM1/2, SW\tablenotemark{c} & 2001-04-14 20:47:25 & 20 \\
\noalign{\smallskip}\tableline\noalign{\smallskip}
\multicolumn{6}{c}{\chandra} \\
\noalign{\smallskip}\tableline\noalign{\smallskip}
5123 & 01:03:56.41 & --72:01:18.7 & ACIS-S & 2003-12-15 04:27:06 & 20 \\
5130 & 01:04:12.31 & --72:01:51.2 & ACIS-S & 2004-04-09 13:08:57 & 19 \\
\enddata
\tablenotetext{a}{Throughout the paper, 
the \fuse\ observations are called N, NE, and SE, as
indicated in the table.}
\tablenotetext{b}{For \fuse\ data.}
\tablenotetext{c}{EPIC-MOS1/2 data in small window mode.}
\end{deluxetable}

\clearpage

\begin{deluxetable}{lcccccc}
\tabletypesize{\footnotesize}
\tablecaption{\label{oneovi}
\fuse\ spectral fit results 
with one set of \ovi\ $\lambda\lambda$1032, 1038
lines per each aperture position. }
\tablehead{
\colhead{Data} & 
\colhead{Bin} & 
\colhead{FWHM} & 
\colhead{Line Shift} & 
\multicolumn{2}{c}{Surface Brightness} &
\colhead{Reduced} \\
\colhead{}& 
\colhead{}&  
\colhead{}&  
\colhead{}& 
\colhead{$F_{\lambda1032}$} & 
\colhead{$F_{\lambda1038}$} & 
\colhead{$\chi^2$} \\
\colhead{}& 
\colhead{[\AA]} & 
\colhead{[km\,s$^{-1}$]} & 
\colhead{[km\,s$^{-1}$]} & 
\colhead{[erg\,cm$^{-2}$\,s$^{-1}$\,arcsec$^{-2}$]} & 
\colhead{[erg\,cm$^{-2}$\,s$^{-1}$\,arcsec$^{-2}$]} & 
\colhead{} }
\startdata
N\tablenotemark{a} & 0.3 & 1540$\pm$350 & 380$\pm$180 & (1.2$\pm0.2) \times 10^{-15}$ & (0.6$\pm0.1) \times 10^{-15}$ & 1.59 \\
NE & 0.3 & 970$\pm$120 & 60$\pm$60 & (1.7$\pm0.3) \times 10^{-15}$ & (1.3$\pm0.4) \times 10^{-15}$ & 1.65 \\
SE & 0.3 & 830$\pm$310 &--160$\pm$30 & (2.3$\pm1.1) \times 10^{-15}$ & (2.0$\pm1.1) \times 10^{-15}$ & 1.72 \\
\enddata
\tablecomments{The flux is de-reddened.
The errors are 90\% confidence ranges.}
\tablenotetext{a}{$\lambda$1038 line flux fixed to half of the $\lambda$1032 flux.}
\end{deluxetable}

\clearpage

\clearpage

\begin{deluxetable}{lccccc}
\tablecaption{\label{ratiotab}
%
%
Line ratios for O, Ne, and Mg. }
\tablehead{
Region & \ovii\ triplet/ & \ovi\ doublet/ & \ovi\ doublet/ & \neix\ triplet/ & 
\mgxi\ triplet/ \\
& \oviii\ Ly$\alpha$ & \ovii\ triplet & \oviii\ Ly$\alpha$ & \nex\ Ly$\alpha$ & 
\mgxii\ Ly$\alpha$ }
\startdata
N & $1.43_{-0.79}^{+0.93}$ & $0.57_{-0.12}^{+0.20}$ & $0.82_{-0.45}^{+0.48}$ &
$1.49_{-0.20}^{+0.26}$ & $4.19_{-2.67}^{+1.85}$ \\
NE & $1.46_{-1.08}^{+1.17}$ & $0.84_{-0.21}^{+0.20}$ & $1.23_{-0.90}^{+0.99}$
& $2.06_{-0.24}^{+0.79}$ & $4.63_{-1.72}^{+3.15}$ \\
SE & $1.08_{-0.32}^{+0.14}$ & $1.04_{-0.38}^{+0.39}$ & $1.12_{-0.52}^{+0.41}$ & 
$2.25_{-0.20}^{+0.37}$ & $7.68_{-6.96}^{+8.73}$ \\
\enddata
\tablecomments{Emission line fluxes are obtained from the
\fuse\ spectra (\ovi\ doublet) and from the \xmmnewton\ spectrum (He-like
triplets and H-like Ly$\alpha$ line of O, Ne, and Mg) by fitting Gaussians.}
\end{deluxetable}

\clearpage

\begin{deluxetable}{lccc}
\tablecaption{\label{xmmpar}
\xmmnewton\ fit spectral results for regions corresponding to \fuse\ MDRS 
aperture positions. }
\tablehead{
 & N & NE & SE }
\startdata
\NH\ (SMC) [$10^{21}$~cm$^{-2}$] \hspace{3.0cm} & 
1.6 (1.3 -- 2.0) & 0.1 (0.0 -- 0.4) &
0.4 (0.2 -- 0.7)
\\
\noalign{\smallskip}\tableline\noalign{\smallskip}
\multicolumn{4}{c}{1.\ VNEI component (O ejecta)} \\
\noalign{\smallskip}\tableline\noalign{\smallskip}
$kT_{1}$ [keV] & 
0.40\tablenotemark{a} & 0.38\tablenotemark{a} &
0.30\tablenotemark{a}
\\
O (solar\tablenotemark{b}) & 
93 (83 -- 104) & 86 (80 -- 94) &
99 (92 -- 107)
\\
$\tau_{1}$ [$10^{10}$~s~cm$^{-3}$] & 
6.0\tablenotemark{a} & 10.\tablenotemark{a} & 
20.\tablenotemark{a}
\\
$K_{1}$\tablenotemark{c} [$10^{-6}$] &
2.4 (2.2 -- 2.6) & 2.3 (1.5 -- 3.2) &
3.7 (3.5 -- 3.9)
\\
\noalign{\smallskip}\tableline\noalign{\smallskip}
\multicolumn{4}{c}{2.\ VNEI component (Ne and Mg ejecta)} \\
\noalign{\smallskip}\tableline\noalign{\smallskip}
$kT_{2}$ [keV] & 
3.5 (2.7 -- 3.7) & 1.5 (1.0 -- 3.2) &
1.1 (0.8 -- 1.6)
\\
Ne (solar\tablenotemark{b}) & 
82 (75 -- 89) & 86 (66 -- 116)&
82 (77 -- 94)
\\
Mg (solar\tablenotemark{b}) & 
24 (20 -- 28) & 33 (23 -- 49) & 
30 (22 -- 40)
\\
$\tau_{2}$ [$10^{10}$~s~cm$^{-3}$] & 
1.8 (1.7 -- 1.9) & 2.3 (1.5 -- 3.7) & 
3.0 (2.1 -- 4.4)
\\
$K_{2}$\tablenotemark{c} [$10^{-6}$] &
= $K_{1}$ & = $K_{1}$ &
= $K_{1}$
\\
\noalign{\smallskip}\tableline\noalign{\smallskip}
\multicolumn{4}{c}{3.\ VNEI component (ISM)} \\
\noalign{\smallskip}\tableline\noalign{\smallskip}
$K_{3}$\tablenotemark{c} [$10^{-4}$] &
3.0 (2.8 -- 3.3) & 2.0 (1.8 -- 2.3) &
2.3 (2.1 -- 2.5)
\\
\noalign{\smallskip}\tableline\noalign{\smallskip}
\multicolumn{4}{c}{Fit information} \\
\noalign{\smallskip}\tableline\noalign{\smallskip}
$\chi^2$ & 559.7 & 608.8 & 618.8
\\
d.o.f & 374 & 368 & 403
\\
Reduced $\chi^2$ & 1.50 & 1.65 & 1.54
\\
\enddata
\tablecomments{The 90\% confidence range is given in parentheses.
The model consists of three VNEI components: one for the blast wave emission,
two for the ejecta emission. 
The absorption for SMC is modeled with SMC 
abundances. Additional Galactic absorption is modeled with fixed column
density of $5.36 \times 10^{20}$~cm$^{-2}$.}
\tablenotetext{a}{
Result from the line diagnostics (Fig.\,\ref{linediagn}).
}
\tablenotetext{b}{
Abundances from \citet{1989GeCoA..53..197A}.
}
\tablenotetext{c}{
Normalization $K = 10^{-14} / (4 \pi D^{2}) \int n_{\rm e} n_{\rm H} dV$, 
where $D$ is the distance to the source (cm), $n_{\rm e}$ is the electron 
density (cm$^{-3}$), and $n_{\rm H}$ is the hydrogen density (cm$^{-3}$).
}
\end{deluxetable}

\clearpage

\begin{deluxetable}{lccc}
\tablecaption{\label{esttab}
Estimated volumes observed through the MDRS apertures, number densities, and 
masses derived from the fits of EPIC-MOS spectra using a three component VNEI 
model.
}
\tablehead{
 & N & NE & SE}
\startdata
%
%
$V$ [pc$^{3}$] \hspace{5.5cm} & 19.8$\pm$4.0 & 24.3$\pm$4.9 & 26.0$\pm$5.2 \\
$V$ [$10^{56}$~cm$^{3}$] & 5.8$\pm$1.2 & 7.1$\pm$1.4 & 7.6$\pm$1.5 \\
%
%
\noalign{\smallskip}\tableline\noalign{\smallskip}
\multicolumn{4}{c}{Number densities} \\
\noalign{\smallskip}\tableline\noalign{\smallskip}
$n_{\rm O} f^{0.5}$~\tablenotemark{c} [cm$^{-3}$]  & 0.045$\pm$0.009 & 0.038$\pm$0.008 & 0.050$\pm$0.009 \\
$n_{\rm Ne} f^{0.5}$ [cm$^{-3}$] & 0.013$\pm$0.005 & 0.012$\pm$0.004 & 0.014$\pm$0.004 \\
$n_{\rm Mg} f^{0.5}$ [cm$^{-3}$] & 0.0012$\pm$0.0004 & 0.0014$\pm$0.0005 & 0.0016$\pm$0.0005 \\
\noalign{\smallskip}\tableline\noalign{\smallskip}
\multicolumn{4}{c}{Mass\tablenotemark{a} assuming a thick shell} \\
\noalign{\smallskip}\tableline\noalign{\smallskip}
$M_{\rm O}/f^{0.5}$ [$M_{\sun}$] & 4.3$\pm$1.6 & 3.7$\pm$1.4 & 4.8$\pm$1.7 \\
$M_{\rm Ne}/f^{0.5}$ [$M_{\sun}$]  & 1.6$\pm$0.8 & 1.4$\pm$0.7 & 1.7$\pm$0.7 \\
$M_{\rm Mg}/f^{0.5}$ [$M_{\sun}$]  & 0.18$\pm$0.08 & 0.21$\pm$0.09 & 0.24$\pm$0.10 \\
\noalign{\smallskip}\tableline\noalign{\smallskip}
\multicolumn{4}{c}{Mass\tablenotemark{a} assuming a thick ring\tablenotemark{b}} \\
\noalign{\smallskip}\tableline\noalign{\smallskip}
$M_{\rm O}/f^{0.5}$  [$M_{\sun}$]  & 4.0$\pm$0.8 & 3.4$\pm$0.7 & 4.4$\pm$0.8 \\
$M_{\rm Ne}/f^{0.5}$ [$M_{\sun}$]  & 1.5$\pm$0.5 & 1.3$\pm$0.4 & 1.6$\pm$0.5  \\
$M_{\rm Mg}/f^{0.5}$ [$M_{\sun}$]  & 0.16$\pm$0.06 & 0.19$\pm$0.06 & 0.22$\pm$0.06 \\
\enddata

\tablenotetext{a}{
O, Ne, and Mg masses in the entire remnant assuming the density in the regions
N, NE, or SE.
}
\tablenotetext{b}{
Assuming a thick ring suggested by \citet{2004ApJ...605..230F}.
}
\tablenotetext{c}{
$f$ is the volume filling factor.
}
\end{deluxetable}

\end{document}